\documentclass[10pt,twocolumn,amsmath,amssymb,amsfonts,aps,prl,showpacs]{revtex4-1}

\usepackage{hyperref}
\usepackage{textcomp}
\usepackage{color}
\usepackage{graphicx}
\usepackage[normalem]{ulem}

\newcommand{\Tr}{\mathop{\rm Tr}\nolimits} %Trace

\begin{document}

\title{Experimental Demonstration of Negative-Valued Polarization Quasi-Probability Distribution}

\author{K.~Yu.~Spasibko$^{1,2,3}$, M.~V.~Chekhova$^{1,2,3}$, F.~Ya.~Khalili$^{2}$}
\affiliation{$^1$Max-Planck-Institute for the Science of Light, G\"unther-Sharowsky-Strasse 1/Bldg. 24, 91058 Erlangen, Germany\\
$^2$Faculty of Physics, M. V. Lomonosov Moscow State University, 119991 Moscow, Russia\\
$^3$Friedrich-Alexander-Universit\"at Erlangen-N\"urnberg, Staudtstrasse 7/B2, 91058 Erlangen, Germany}


\begin{abstract}

Polarization quasiprobability distribution defined in the Stokes space shares many important properties with the Wigner function for the position and momentum. Most notably, they both give correct one-dimensional marginal probability distributions and therefore represent the natural choice for the probability distributions in classical hidden-variable models. In this context, negativity of the Wigner function is considered as a proof of non-classicality for a quantum state. On the contrary, the polarization quasiprobability distribution demonstrates negativity for all quantum states. This feature comes from the discrete nature of the Stokes variables; however, it was not observed in previous experiments, because they were performed with photon-number averaging detectors. Here we reconstruct the polarization quasiprobability distribution of a coherent state with photon-number resolving detectors, which allows us to directly observe for the first time its negativity.

\end{abstract}

\maketitle

\paragraph{Introduction.}

Non-commuting observables are nonexistent in classical physics, but arise in quantum mechanics and optics. They lead to the difficulties in the attempts to describe quantum states in a semiclassical way, because it is impossible to define a joint probability distribution for such observables. As a remedy for this, {\it quasi}probability distributions were proposed, which can take negative values and therefore violate one of the main axioms of the probability theory.

The most well-known example of non-commuting observables is the canonical pair of position and momentum and the most remarkable corresponding joint quasiprobability distribution is the Wigner one~\cite{Wigner1932}. Its major distinctive feature is that, in contrast to {\it e.g.} the Glauber-Sudarshan $P$-representation~\cite{Glauber1963,Sudarshan1963} or the Husimi-Kano $Q$-representation~\cite{Husimi1940,Kano1965}, it gives correct marginal distributions for the position and momentum \cite{Schleich2001}. Therefore, it represents the natural choice for the probability distributions in the classical hidden variables models. Because of this property, it is widely accepted that the negativity of a Wigner distribution means the non-classicality of the quantum state \cite{Hudson1974,Soto1983,Schleich2001,Lvovsky2001,Raymer2004,Chekhova2013}.

Due to the unique features of the Wigner function, mathematical objects with similar properties were defined for many different systems and observables. In particular, it was done for the discrete-valued position and momentum \cite{Bjork2008}, for the  Hermite-Gaussian and Laguerre-Gaussian modes of an optical beam \cite{Simon2000}, and for the canonical pair of the angle and the angular momentum of vortex states \cite{Rigas2008}.

The analog of the Wigner distribution for the three non-commuting Stokes observables [see Eqs. (\ref{Stokes_intro})], the polarization quasiprobability distribution (PQPD), was developed in Refs.~\cite{Bushev2001, Karassiov2002}. PQPD gives correct one-dimensional marginal probability distributions for all Stokes observables and their linear combinations. A very interesting feature of this distribution is that it takes negative values for all quantum states of light, even for the ``most classical'' coherent ones. The physical origin of this behaviour was explored theoretically in Ref.~\cite{Chekhova2013}. The negativity was shown to appear because the Stokes observables are discrete-valued. At the same time, this feature was never observed in polarization tomography experiments, see {\it e.g.} \cite{Bushev2001, Marquardt2007, Agafonov2012, Kanseri2012, Mueller2012}, because all these experiments were performed with photon-number averaging detectors, which smoothed the measured photon-number statistics and washed out the non-classical features of PQPD.

In this work, we have measured PQPD for a coherent state of light using, for the first time to the best of our knowledge, single-photon detectors.  We have developed the reconstruction procedure for this case, which allowed us to restore the PQPD with a high quality using a limited data set. The reconstructed distribution demonstrates well-pronounced negative-valued areas.

\paragraph{Stokes observables and PQPD.}

A quantum state of light can be fully described by its density operator $\hat{\rho}$. The PQPD $W(S_1,S_2,S_3)$ for such a state is defined as the Fourier transform of the polarization characteristic function $\chi(u_1,u_2,u_3)$,
\begin{eqnarray}
W(S_1,S_2,S_3)&=&\int_{-\infty}^{\infty}\chi(u_1,u_2,u_3)\nonumber\\
&\times&\exp\left(-i\sum_{i=1}^3u_i S_i\right)\frac{du_1du_2du_3}{(2\pi)^3},
\label{W_basis}
\end{eqnarray}
where
\begin{equation}
\chi(u_1,u_2,u_3)=\Tr\left[\hat{\rho}\exp\left(i\sum_{i=1}^3u_i \hat{S}_i\right)\right].
\end{equation}
The Stokes operators $\hat{S}_i$ are defined as
\begin{equation}
\begin{array}{c}
\hat{S}_1=\hat{n}_H-\hat{n}_V,\qquad \hat{S}_2=\hat{a}_V^\dagger\hat{a}_H+\hat{a}_H^\dagger\hat{a}_V,\\
\hat{S}_3=i(\hat{a}_V^\dagger\hat{a}_H-\hat{a}_H^\dagger\hat{a}_V),
\end{array}
\label{Stokes_intro}
\end{equation}
%\begin{equation}
%\begin{array}{ll}
%\hat{S}_0=\hat{n}_H+\hat{n}_V, & \hat{S}_1=\hat{n}_H-\hat{n}_V,\\
%\hat{S}_2=\hat{a}_V^\dagger\hat{a}_H+\hat{a}_H^\dagger\hat{a}_V, & \hat{S}_3=i(\hat{a}_V^\dagger\hat{a}_H-\hat{a}_H^\dagger\hat{a}_V),
%\end{array}
%\label{Stokes_intro}
%\end{equation}
where $\hat{a}_H$ and $\hat{a}_V$ are the photon annihilation operators for the horizontal (H) and vertical (V) polarization modes, $\hat{n}_{H,V}=\hat{a}_{H,V}^\dagger\hat{a}_{H,V}$ are photon-number operators in these modes. All Stokes operators can be represented as the differences of photon-number operators in certain modes, therefore the corresponding Stokes observables (e.g. $S_{1}$) can only take integer values $n\in\mathbb{Z}$.

%For exploring negativity of polarization quasiprobability distribution we used a linearly polarized weak coherent state. If we assume that our state is horizontally polarized then density matrix could be written in the following way:
%\begin{equation}
%\hat{\rho}=p_0
%\end{equation}

%One can also introduce generalized Stokes operator $\hat{S}_{\alpha\beta}$
%\begin{equation}
% =(\hat{S}_1\cos\alpha+\hat{S}_2\sin\alpha)\cos\beta+\hat{S}_3\sin\beta,
%\end{equation}
%where angles $\alpha\in[0, 2\pi]$ and $\beta\in[-\pi/2, \pi/2]$ denote point on Poincar\'e sphere in Stokes space.

%In this notation Stokes parameters $S_i$ are defines as mean values of Stokes operators $\left<\hat{S}_i\right>$.
%and normalized Stokes parameters $s_j$ are defines as
%\begin{equation}
%s_j=S_i/S_0,\quad j=1,2,3.
%\end{equation}

%\section{Experiment}
\paragraph{PQPD reconstruction.}

\begin{figure}
\includegraphics[width=8.5cm]{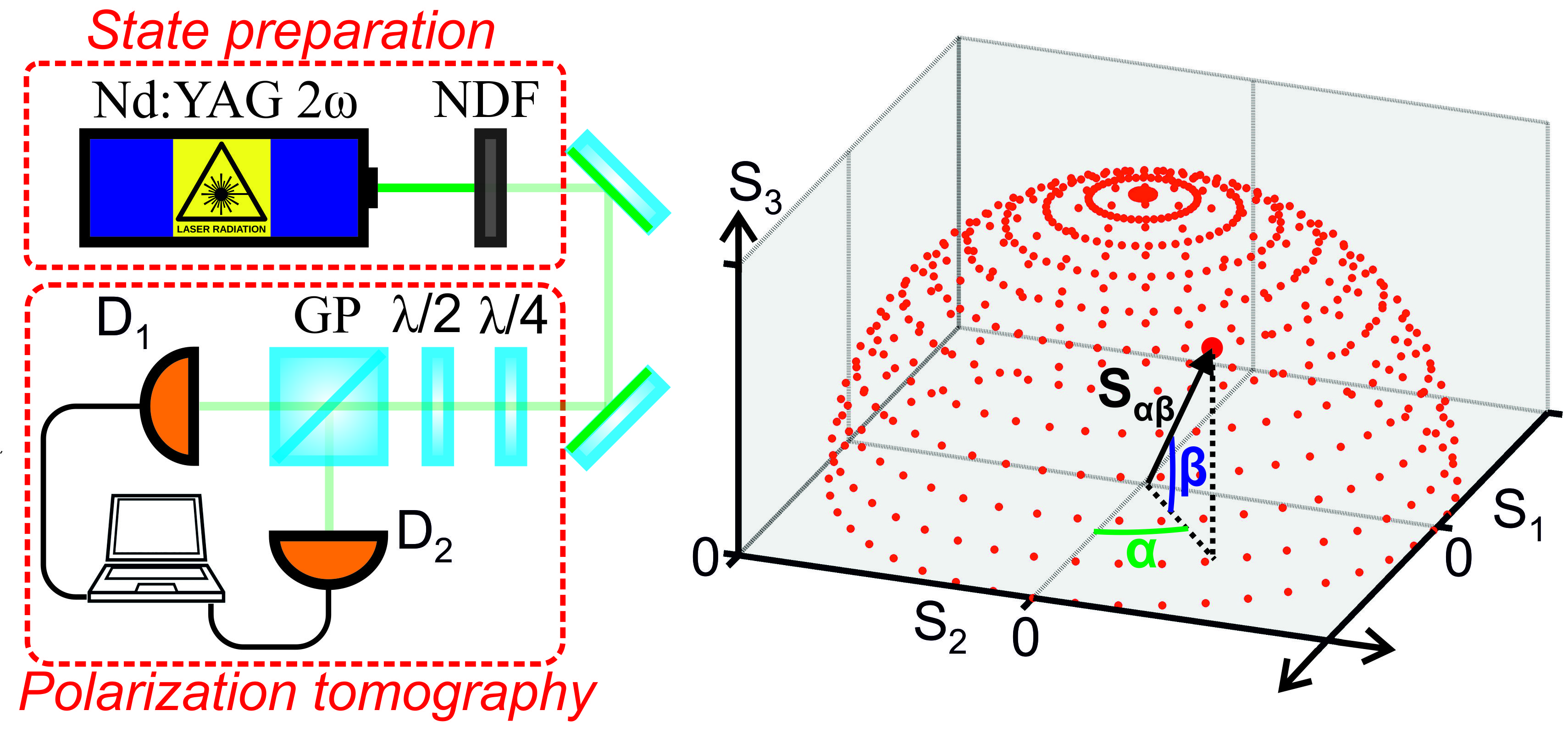}
\caption{Left: experimental setup. A weak coherent state is prepared by attenuating the second harmonic of a Nd:YAG laser (Nd:YAG~2$\omega$) with neutral density filters (NDF). A standard setup for polarization tomography consists of a quarter- and a half-wave plates ($\lambda/4$ and $\lambda/2$), a polarizing beam splitter, and two detectors (D$_1$ and D$_2$). We use a Glan-Taylor prism (GP) as a polarizing beam splitter and two avalanche photodiodes as detectors. Right: the points at which tomographic measurements are performed are shown on the Poincar\'e sphere.}
\label{Setup}
\end{figure}

A standard setup for polarization tomography (see Fig.~\ref{Setup}) consists of a quarter- and a half-wave plates ($\lambda/4$ and $\lambda/2$), a polarizing beam splitter and two detectors (D$_1$ and D$_2$). For each pair of settings of the quarter-wave ($\tilde{\beta}$) and half-wave ($\tilde{\alpha}$) plates, such a setup measures a different arbitrary Stokes operator $\hat{S}_{\alpha\beta}=\hat{n}_{1}-\hat{n}_{2}$. The operators $\hat{n}_{1,2}$ correspond to the photon numbers in the mode transmitted or reflected by the polarizing beam splitter and are measured by D$_{1}$ or D$_{2}$, respectively.

The angles $\alpha\in[0,2\pi]$ and $\beta\in[-\pi/2,\pi/2]$ that define a point on the Poincar\'e sphere (see Fig.~\ref{Setup}) are determined by the settings of the wave plates,
\begin{equation}
\alpha=4\tilde{\alpha}-2\tilde{\beta},\quad\beta=2\tilde{\beta}.
\end{equation}

An arbitrary Stokes operator $\hat{S}_{\alpha\beta}$ can be represented in Cartesian coordinates $(\hat{S}_1,\hat{S}_2,\hat{S}_3)$ as
\begin{equation}
\hat{S}_{\alpha\beta}=(\hat{S}_1\cos\alpha+\hat{S}_2\sin\alpha)\cos\beta+\hat{S}_3\sin\beta.
\end{equation}
It is clear that this operator possesses inversion symmetry $\hat{S}_{(\alpha+\pi)(-\beta)}=-\hat{S}_{\alpha\beta}$, thus measurements only on the half of the Poincar\'e sphere suffice for the full reconstruction of any state.

In the experiment, for each point on the Poincar\'e sphere (for each $\alpha$ and $\beta$), acquisition of many $S_{\alpha\beta}$ values is needed. From these values we calculate the probabilities $W_{\alpha\beta}(n)$ that $S_{\alpha\beta}$ are equal to $n$.

From these probabilities we restore the polarization characteristic function $\chi_{\alpha\beta}(\lambda)$ in spherical coordinates  ($\lambda,\alpha,\beta$) \cite{Chekhova2013}:
\begin{equation}
\chi_{\alpha\beta}(\lambda)=\sum_{n=-\infty}^{\infty}W_{\alpha\beta}(n)e^{i\lambda n},\quad\lambda\in[0,\infty).
\end{equation}
These spherical coordinates ($\lambda,\alpha,\beta$)  are related to the Cartesian ones ($u_1,u_2,u_3$) by the following transformations:
\begin{equation}
\begin{array}{c}
u_1=\lambda\cos\alpha\cos\beta,\qquad u_2=\lambda\sin\alpha\cos\beta,\\
u_3=\lambda\sin\beta.
\end{array}
\end{equation}

Thus, using these transformations, Eq. (\ref{W_basis}) can be rewritten as
%noticing that $W_{\alpha\beta}(n)=W_{(\alpha+\pi)(-\beta)}(-n)$ (comes from the fact that $W(S_1,S_2,S_3)\in\mathbb{R}$) for the bottom part of Poincar\'e sphere ($\beta<0$)
%\begin{equation}
%\begin{array}{c}
%u_1=\lambda\cos\alpha\cos\beta,\qquad u_2=\lambda\sin\alpha\cos\beta,\\
%u_3=\lambda\sin\beta,
%\end{array}
%\end{equation}
%taking into account that $\chi(u_1,u_2,u_3)=\chi_{\alpha\beta}(\lambda)$ and noticing that $W_{\alpha\beta}(n)=W_{(\alpha+\pi)(-\beta)}(-n)$ (comes from the fact that $W(S_1,S_2,S_3)\in\mathbb{R}$) for the bottom part of Poincar\'e sphere ($\beta<0$), and finally performing integration over $\lambda$ we obtained equation for the reconstruction of
\begin{eqnarray}
W(S_1,S_2,S_3)&=&-\frac{1}{(2\pi)^2}\int_0^{2\pi}d\alpha\int_0^{\pi/2}d\beta\cos\beta\nonumber\\
&\times&\sum_{n=-\infty}^{\infty}W_{\alpha\beta}(n)\delta^{(2)}(S_{\alpha\beta}-n),
\label{W_exp_eq}
\end{eqnarray}
where $\delta^{(2)}(x)$ is the second derivative of the Dirac delta function. Here we exploit the symmetry of $\hat{S}_{\alpha\beta}$ and perform integration over the radial coordinate $\lambda$. As a result, we obtain the equation for reconstructing the PQPD $W(S_1,S_2,S_3)$ from the experimentally measured probabilities $W_{\alpha\beta}(n)$.

The reconstruction of PQPD $W_\epsilon(S_1,S_2,S_3)$ from the experimentally acquired data set using Eq.\,\eqref{W_exp_eq} requires some approximation $\delta_\epsilon(x)$ for the Dirac delta function $\delta(x)$, where $\epsilon$ is the smoothing parameter. We choose the Gaussian approximation,
\begin{equation}
\delta_\epsilon(x)=\frac{1}{2\epsilon\sqrt{\pi}}e^{-x^2/4\epsilon^2},
\label{Gauss_app}
\end{equation}
and similarly for the derivatives of $\delta(x)$. The smoothing parameter $\epsilon$ should be chosen from the following considerations. On the one hand, it has to be small enough to represent all features of the PQPD, but on the other hand, small values of $\epsilon$ lead to a lot of artifacts in the reconstructed distribution (the so-called reconstruction noise).
%On the one hand this approximation is quite simple and on the other hand it gives good suppression of the reconstruction noise (e.g. better than Lorentzian approximation).

%For the experimental reconstruction $\epsilon$ was chosen equal to 0.02. This value was a good compromise between the quality of the reconstruction (large values of $\epsilon$ smooth features of the distribution) and the reconstruction noise (the smaller $\epsilon$ the bigger the noise).

\paragraph{Experiment and data processing.}

We have performed the polarization tomography of a horizontally polarized weak coherent state $\left|\gamma\right>$. This state was produced by strongly attenuating a coherent beam at the wavelength 532~nm generated by a pulsed Nd:YAG~laser (Nd:YAG~2$\omega$) with the pulse duration 10 ns and repetition rate 10 kHz (see Fig.~\ref{Setup}). Attenuation (or any other linear losses) does not change the statistical properties of a coherent state: the state remains coherent, but the mean number of photons $|\gamma|^2$ is reduced. The attenuation to a single-photon level was performed by a neutral density filter (NDF). It was done in such a way that the probability of single-photon detection events $p_1\approx|\gamma|^2$ was equal to 0.189. In this case $p_1$ was at least one order of magnitude bigger than the probabilities of two-photon and higher-order detection events. Therefore we ignored such events and considered only single-photon and no-photon detection events (with the probability $p_0$). We used avalanche photodiodes as single-photon detectors (D$_1$ and D$_2$).

%The tomography was performed by means of quarter- and half-wave plates ($\lambda/4$ and $\lambda/2$), Glan-Taylor prism polarizer and two avalanche photodiodes (APD). This setup measures Stokes operator $\hat{S}_{\alpha\beta}=\hat{n}_{1}-\hat{n}_{2}$, where $\hat{n}_{1,2}$ is photon-operator in transmitted or reflected (by GP) mode and measured by D$_{1}$ or D$_{2}$ respectively. Angles on Poincar\'e sphere $\alpha$ and $\beta$ were calculated from settings for the quarter- ($\tilde{\beta}$) and half-wave plates ($\tilde{\alpha}$) with the following equations
%\begin{equation}
%\alpha=4\tilde{\alpha}-2\tilde{\beta},\quad\beta=2\tilde{\beta}.
%\end{equation}

%\paragraph{Data processing.}

The points $(\alpha_k,\beta_l)$ on the Poincar\'e sphere where tomographic measurements were performed cover the upper hemisphere ($\beta\ge0$) with a step of $8^\circ$ degrees (see Fig.~\ref{Setup}). These points have been accessed by different combinations of the settings for the quarter- and half-wave plates with the steps equal to $4^\circ$ and $2^\circ$ degrees, respectively (and for $\tilde{\beta}=45^\circ$, the `north' pole of the Poincar\'e sphere was accessed). For each point from this discrete set we have calculated the experimental probabilities $\tilde{W}_{\alpha_k\beta_l}(n)$, where $n=\{-1,0,1\}$.

The full experimental dataset ${\tilde{W}_{\alpha_k\beta_l}(n)}$ is not suitable for the final integration over $\alpha$ and $\beta$ in Eq. (\ref{W_exp_eq}), because it is defined on a discrete set $\{\alpha_k,\beta_l$\}. Thus it should be interpolated by a continuous function. The interpolated function $W_{\alpha\beta}(n)$ is given by the convolution sum of the data points $\tilde{W}_{\alpha_k\beta_l}(n)$ with the interpolation kernel $u(\alpha,\beta)$,
\begin{equation}
W_{\alpha\beta}(n)=\sum_{\alpha_k,\beta_l}\tilde{W}_{\alpha_k\beta_l}(n)u(\alpha-\alpha_k,\beta-\beta_l).
\end{equation}

Various interpolation kernels can be used. The simplest one is a rectangular function $u(\alpha,\beta)=\Pi(\alpha)\Pi(\beta)$, where
\begin{equation}
\Pi(x) = \left\{
\begin{array}{rl}
1,& |x|<1/2\\
0,& |x|\ge1/2.\\
\end{array}
\right.
\end{equation}

The integration of thus interpolated function (e.g. as part of the Fourier or Radon transform) gives exactly the same result as when the integration is replaced by the summation. Such a replacement was always used for the reconstruction in the polarization tomography \cite{Marquardt2007,Agafonov2012,Kanseri2012,Mueller2012}. Unfortunately, with this interpolation, the transformations are accompanied by rather high noise. One can overcome this problem by collecting more experimental points $(\alpha_k,\beta_l)$ or by using different interpolation kernels.

Interpolation methods are well-developed for image resampling \cite{Maeland1988, Parker1983}. It has been shown that several interpolation kernels could suppress the reconstruction noise by more than 30 dB better than the rectangular-function kernel.

In our case the probabilities $W_{\alpha\beta}(n)$ could not be negative; hence we needed a strictly positive kernel. We chose to use a positive cubic spline kernel $u(\alpha,\beta)=u(\alpha)u(\beta)$ \cite{Maeland1988}, where
\begin{equation}
u(x) = \left\{
\begin{array}{cl}
2|x|^3-3|x|^2+1,& |x|\le1,\\
0,& |x|>1.\\
\end{array}
\right.
\end{equation}
This kernel suppresses the noise very well and is at the same time quite simple. For each interval between the data points, e.g. $(x_k,x_{k+1})$, the interpolation requires only the experimental data from the endpoints of the interval ($x_k$ and $x_{k+1}$). Hence this kernel has the same simplicity as the linear interpolation kernel, but a better performance.

%because $W_{\alpha\beta}$ is continuous function interpolation of discrete experimental dataset $\tilde{W}_{\alpha_k\beta_l}(n)$ measured at points $(\alpha_k,\beta_l)$ were used (see Appendix for the details).

%As we mentioned in our experiment we consider only single-photon events and no-photon events so for the single-shot measurement generalized Stokes parameter $S_{\alpha\beta}$ could be equal only to the $n=\{1,0,-1\}$ and for each point on Poincar\'e sphere (for each $\alpha$ and $\beta$) we get probabilities $W_{\alpha\beta}(n)$.

%\textcolor{magenta}{move here important parts from Appendix B}

%The reconstruction is possible only for the smoothed quasiprobability distribution $W(S_1,S_2,S_3)$ when some approximation for the $\delta^{(2)}$ is used. Moreover, because $W_{\alpha\beta}$ is continuous function interpolation of discrete experimental dataset $\tilde{W}_{\alpha_k\beta_l}(n)$ measured at points $(\alpha_k,\beta_l)$ were used (see Appendix for the details). \textcolor{magenta}{Rewrite this}

%\section{Results}

\begin{figure}
\includegraphics[width=8.5cm]{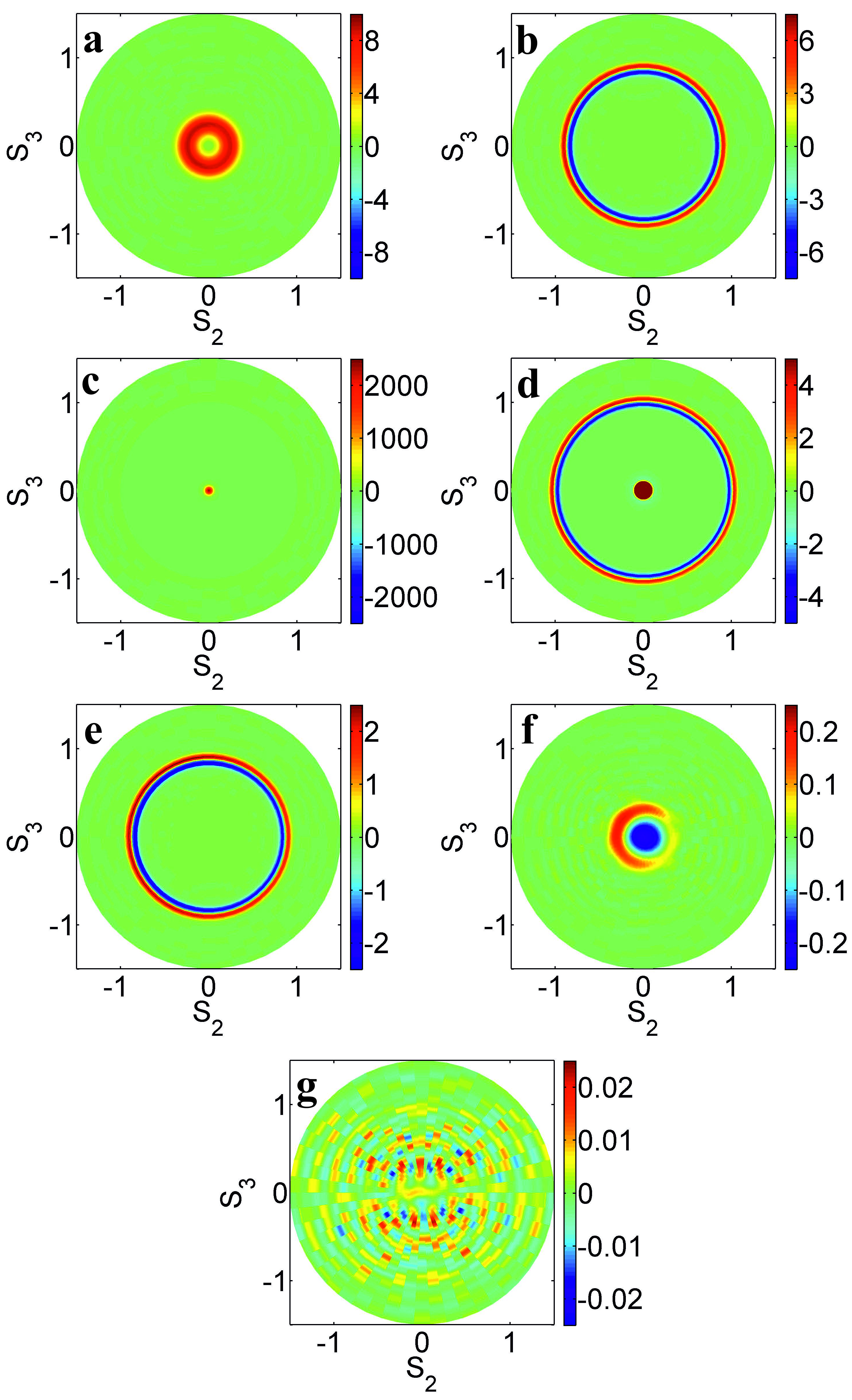}
\caption{Cross-sections of the reconstructed PQPD $W_\epsilon(S_1,S_2,S_3)$ (with $\epsilon=0.02$) along the $(S_2,S_3)$ plane at $S_1=1$ (a), $S_1=0.5$ (b), $S_1=0$ (c,d), $S_1=-0.5$ (e), $S_1=-1$ (f) and $S_1=-1.5$ (g). In panel (d), the same color is used for values larger than 5 to highlight the jump at $S=1$.}
\label{Exp_S1}
\end{figure}

\begin{figure}
\includegraphics[width=8.5cm]{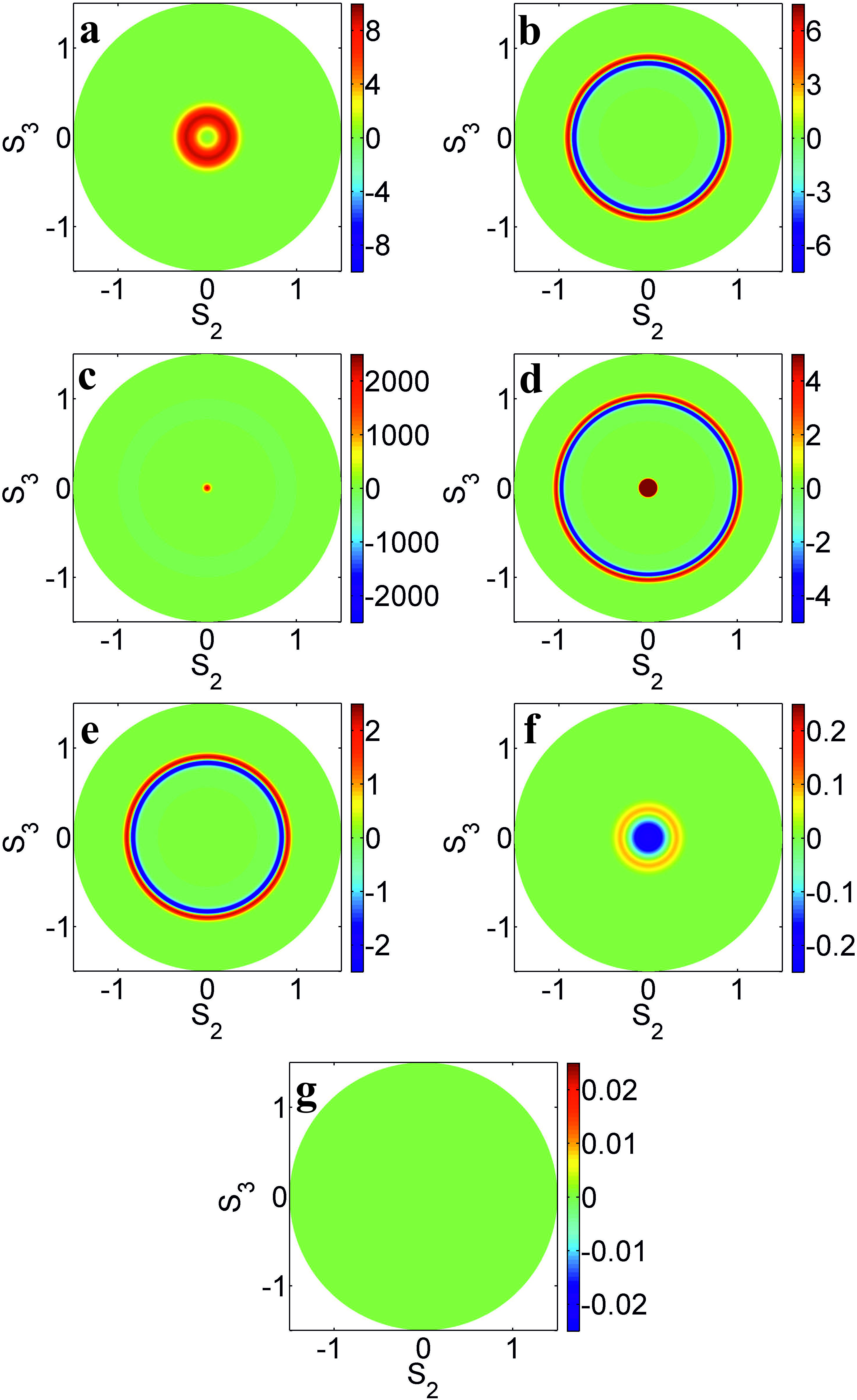}
\caption{Cross-sections of the theoretical PQPD $W_\epsilon(S_1,S_2,S_3)$ smoothed by $\epsilon=0.02$ along $(S_2,S_3)$ plane at $S_1=1$ (a), $S_1=0.5$ (b), $S_1=0$ (c,d), $S_1=-0.5$ (e), $S_1=-1$ (f) and $S_1=-1.5$ (g). In panel (d), the same color is used for values larger than 5 to highlight the jump at $S=1$.}
\label{The_S1}
\end{figure}

\paragraph{Results.}

Using this interpolation and the approximation (\ref{Gauss_app}) with $\epsilon=0.02$, we have reconstructed the PQPD $W_\epsilon(S_1,S_2,S_3)$. Its cross-sections along the $(S_2,S_3)$ plane at different values of $S_1$ are shown in  Fig.~\ref{Exp_S1}.

%The cross-sections of the reconstructed quasiprobability distribution clearly show that for the coherent state distribution has negative values.

In general, each distribution contains a central peak at the origin of the Stokes space ($S=\sqrt{S_1^2+S_2^2+S_3^2}=0$) and a jump from negative values to positive ones at $S=1$. The central peak, which appears because of the no-photon detection events, is more than two orders of magnitude higher than the jump, which happens because of the single-photon ones. At values $S>1$ there is only the reconstruction noise (Fig.~\ref{Exp_S1}g).

The reconstructed distribution $W_\epsilon(S_1,S_2,S_3)$ is in agreement with the theoretical one that is derived for our case (single-photon and no-photon detection events) in spherical coordinates $(S,\theta,\phi)$ \cite{SeeSupp}:

\begin{eqnarray}
W(S,\theta,\phi)&=&p_0\delta_3(S)+\frac{p_1\cos\theta}{4\pi S^2}\delta(S-1)\nonumber\\
&-&\frac{p_1(1+\cos\theta)}{4\pi S}\delta'(S-1),
\label{Theor_W}
\end{eqnarray}
where $\delta_3(S) = \delta(S_1)\delta(S_2)\delta(S_3)$, $\delta'(x)$ is the first derivative of the Dirac delta function, and
\begin{equation}
\begin{array}{c}
S_1=S\cos\theta,\qquad S_2=S\sin\theta\cos\phi,\\
S_3=S\sin\theta\sin\phi.
\label{S_cart_sph}
\end{array}
\end{equation}

From these formulas we have calculated the theoretical PQPD $W_\epsilon(S_1,S_2,S_3)$ for the same probabilities of single-photon ($p_1=0.189$) and no-photon detection events ($p_0=0.811$) as in the experimental case. We used the same approximation (\ref{Gauss_app}) and the same value of the smoothing parameter $\epsilon=0.02$. The same cross-sections are shown for both distributions (Fig.~\ref{The_S1}). The experimental and theoretical distributions are almost indistinguishable. The only differences are caused by the reconstruction noise (Fig.~\ref{Exp_S1}g) and imperfections of the half- and quarter-wave plates (Fig.~\ref{Exp_S1}f).

\begin{figure}
\includegraphics[width=8.5cm]{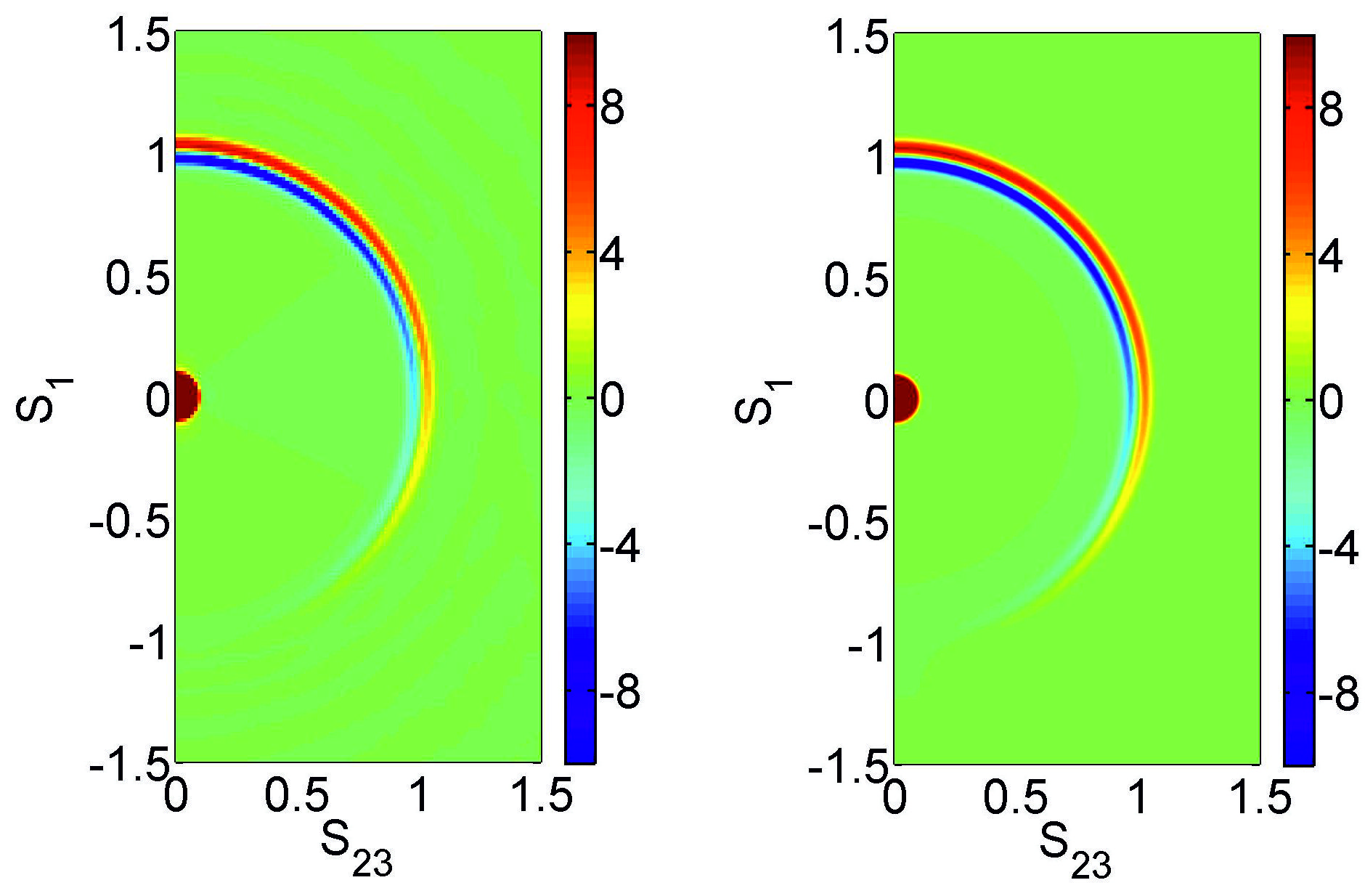}
\caption{Cross-sections of the experimental (left) and theoretical (right) PQPD $W_\epsilon(S_1,S_{23},\phi)$ (with $\epsilon=0.02$) at $\phi=0$. In all figures, the same color is used for values larger than 10 to highlight the jump at $S=1$.}
\label{Exp_The_Phi}
\end{figure}

It is clear that the distribution $W_\epsilon(S_1,S_2,S_3)$ possesses a rotation symmetry in the plane $(S_2,S_3)$. Thus it is convenient to use cylindrical coordinates $(S_1,S_{23},\phi)$, with the radial coordinate $S_{23}=\sqrt{S_2^2+S_3^2}=S\sin\theta$, instead of the Cartesian ones $(S_1,S_2,S_3)$.
%\begin{equation}
%S_2=S_{23}\cos(\phi),\qquad S_3=S_{23}\sin(\phi).
%\end{equation}
Due to this symmetry, up to experimental imperfections a cross-section at some angle $\phi$ (e.g. $\phi=0$) presents all features of the PQPD (Fig.~\ref{Exp_The_Phi}).

\paragraph{Conclusion.}

We have shown experimentally the full reconstruction of PQPD with photon-number resolving detectors. As a result we observed the intrinsic negativity of PQPD originating from the discrete nature of the Stokes observables. The last feature has been never observed before because previous experiments were realized with photon-number averaging detectors. For our reconstruction we have elaborated a procedure that leads to high-quality PQPD from a relatively small dataset. The PQPD reconstruction with photon-number resolving detectors is very promising because of novel detectors of this kind that can resolve up to tens of photons with more than 90\% quantum efficiency \cite{Fukuda2011,Miki2014,Allman2015}. These detectors can push forward this direction in the polarization tomography and make it a useful tool for quantum state characterization.

%We have shown experimentally that, despite a common opinion, the negativity of \textcolor{blue}{certain quasiprobability distributions} does not exactly mean the non-classicality of the state. In the polarization tomography the quasiprobablity distribution can have negative values even for the coherent state. A similar result was recently shown experimentally for the Wigner distribution defined for the Laguerre-Gaussian modes of an optical beam \cite{Stoklasa2015}.

%This happens because of photon-number resolving detection is intrinsically non-Gaussian. We also developed and performed in experiment a method for reconstructing polarization quasiprobability distribution from measurements with photon-number resolving detection.

%\section{Acknowledgments}

We acknowledge the financial support of the Russian Foundation for Basic Research grants 14-02-31030 and 14-02-00399. The work of F.~Ya.~Khalili was supported by LIGO NSF grant PHY-1305863.

\end{document}

\title{Supplementary Material: Experimental Demonstration of Negative-Valued Polarization Quasiprobability Distribution}
\author{K.~Yu.~Spasibko$^{1,2,3}$, M.~V.~Chekhova$^{1,2,3}$, F.~Ya.~Khalili$^{2}$}
\affiliation{$^1$Max-Planck-Institute for the Science of Light, G\"unther-Sharowsky-Strasse 1/Bldg. 24, 91058 Erlangen, Germany\\
$^2$Faculty of Physics, M. V. Lomonosov Moscow State University, 119991 Moscow, Russia\\
$^3$Friedrich-Alexander-Universit\"at Erlangen-N\"urnberg, Staudtstrasse 7/B2, 91058 Erlangen, Germany}

\maketitle

\section{Calculation of the theoretical PQPD}

To describe the reconstructed PQPD $W(S_1,S_2,S_3)$ we calculate the theoretical distribution for our case. Thus we consider a horizontally polarized weak coherent state and only single-photon and no-photon detection events. In this case the density operator looks like
$$
\hat{\rho} = \left[p_0\ket{0}\bra{0}_{H}+p_1\ket{1}\bra{1}_H\right]\otimes\ket{0}\bra{0}_V.
\refstepcounter{suppeq}
\eqno{(\rm S\arabic{suppeq})}
\label{rho}
$$
Similarly to the case of the experimental data processing, the starting point is the definition of the PQPD and of the  polarization characteristic function [see Eqs. (1) and (2)].

Derivation of the theoretical distribution requires the use of spherical coordinates $(\lambda,\xi,\rho)$ that differ from $(\lambda,\alpha,\beta)$ used in the experimental data processing:
$$
\begin{array}{c}
u_1=\lambda\cos\xi,\qquad u_2=\lambda\sin\xi\cos\rho,\\
u_3=\lambda\sin\xi\sin\rho.
\end{array}
\refstepcounter{suppeq}
\eqno{(\rm S\arabic{suppeq})}
$$

The polarization characteristic function for the density operator (\ref{rho}) consists of the no-photon and single-photon parts~\cite{Chekhova2013}:
$$
\chi_{\xi\rho}(\lambda) = p_0 + p_1(\cos\lambda + i\sin\lambda\cos\xi).
\refstepcounter{suppeq}
\eqno{(\rm S\arabic{suppeq})}
$$

The no-photon part of the characteristic function can be directly integrated in Eq. (1). The resulting no-photon PQPD $W_0$ is a $\delta$-peak at the origin of the Stokes space,
$$
W_0(S_1,S_2,S_3) = p_0\delta(S_1)\delta(S_2)\delta(S_3).
\refstepcounter{suppeq}
\eqno{(\rm S\arabic{suppeq})}
\label{W_0}
$$

Derivation of a single-photon PQPD $W_1$ is more tricky. As in the derivation of Eq. (8), we pass from the Cartesian coordinates $(u_1,u_2,u_3)$ to the spherical ones $(\lambda,\xi,\rho)$ and perform the integration over $\lambda$. Then the single-photon distribution looks like

$$
\begin{array}{l}
W_1(S_1,S_2,S_3)=\\[3pt]
\displaystyle-\frac{p_1}{2(2\pi)^2}\int_0^{2\pi}d\rho\int_0^{\pi}d\xi\sin\xi(1+\cos\xi)\delta^{(2)}(S_{\xi\rho}-1),
\end{array}
\refstepcounter{suppeq}
\eqno{(\rm S\arabic{suppeq})}
\label{W1_1}
$$
where an arbitrary Stokes observable $S_{\xi\rho}$ is defined as
$$
S_{\xi\rho}=S_1\cos\xi+(S_2\cos\rho+S_3\sin\rho)\sin\xi.
\refstepcounter{suppeq}
\eqno{(\rm S\arabic{suppeq})}
$$

%Using Leibniz integral rule formula (\ref{W1_1}) was transformed in the following form
%$$
%\begin{array}{l}
%\displaystyle W_1(S_1,S_2,S_3)=-\frac{p_1}{2(2\pi)^2}\times\\[3pt]
%\vphantom{.}\\
%\displaystyle\left.\frac{\partial^2}{(\partial x)^2}\left[\int_0^{2\pi}d\rho\int_0^{\pi}d\xi\sin\xi(1+\cos\xi)\delta(S_{\xi\rho}-x)\right]\right|_{x=1},
%\end{array}
%\refstepcounter{suppeq}
%\eqno{(\rm S\arabic{suppeq})}
%$$

It is possible to perform the integration over $\rho$ in Eq. (\ref{W1_1}) using the spherical coordinates $(S,\theta,\phi)$ defined in Eqs. (14). In this case
$$
S_{\xi\rho}=S(\cos\xi\cos\theta+\sin\xi\sin\theta\cos\bar{\rho}),
\refstepcounter{suppeq}
\eqno{(\rm S\arabic{suppeq})}
$$
where $\bar{\rho}=\rho-\phi$. Using the Leibniz integral rule, Eq. (\ref{W1_1}) is transformed into
$$
W_1(S,\theta,\phi)=-\frac{p_1}{2(2\pi)^2}\left.\frac{\partial^2I_\xi}{(\partial y)^2}\right|_{y=1},
\refstepcounter{suppeq}
\eqno{(\rm S\arabic{suppeq})}
\label{W1_2}
$$
where
$$
I_\xi=\int_0^{\pi}d\xi\,\frac{1+\cos\xi}{S\sin\theta}I_{\bar\rho},
\refstepcounter{suppeq}
\eqno{(\rm S\arabic{suppeq})}
\label{I_xi}
$$
$$
I_{\bar\rho}=\lim_{\kappa\to0}\int_0^{2\pi}d\bar\rho\,\delta_\kappa(P-\cos\bar\rho),
\refstepcounter{suppeq}
\eqno{(\rm S\arabic{suppeq})}
$$
and
$$
P=\frac{y-S\cos\xi\cos\theta}{S\sin\xi\sin\theta}.
\refstepcounter{suppeq}
\eqno{(\rm S\arabic{suppeq})}
$$
Here we use the rectangular approximation [see Eq. (11)] for the Dirac delta function
$$
\delta_\kappa(x)=\frac{1}{\kappa}\Pi\left(\frac{x}{\kappa}\right).
\refstepcounter{suppeq}
\eqno{(\rm S\arabic{suppeq})}
$$

%Using this approximation integral $\tilde{I}_{\bar\rho}$ became quite simple to integrate
%$$
%\begin{array}{l}
%\tilde{I}_{\bar\rho} = \\[3pt]
%\frac{2}{\kappa}\left\{
%\begin{array}{cl}
%0,&|P|>1+\frac{\kappa}{2},\\[3pt]
%\arccos\left(|P|-\frac{\kappa}{2}\right),&||P|-1|<\frac{\kappa}{2},\\[3pt]
%\arccos\left(P-\frac{\kappa}{2}\right)-\arccos\left(P+\frac{\kappa}{2}\right),&|P|<1-\frac{\kappa}{2}.
%\end{array}
%\right.
%\end{array}
%\refstepcounter{suppeq}
%\eqno{(\rm S\arabic{suppeq})}
%$$
After integrating over $\bar\rho$ and simplifying the obtained result by setting $\kappa\to0$, we achieve
$$
I_{\bar\rho}=\lim_{\kappa\to0}\left\{
\begin{array}{cl}
\displaystyle\frac{2}{\sqrt{1-P^2}},&|P|<1-\frac{\kappa}{2},\\[3pt]
\frac{2\sqrt{2}}{\kappa}\sqrt{1-|P|+\frac{\kappa}{2}},&||P|-1|\le\frac{\kappa}{2},\\[3pt]
0,&|P|>1+\frac{\kappa}{2}.
\end{array}
\right.
\refstepcounter{suppeq}
\eqno{(\rm S\arabic{suppeq})}
$$

Note that the second case leads to a singularity $\propto1/\sqrt{\kappa}$. However, because the width of the corresponding integration segment is proportional to $\kappa$, it gives no contribution into the integral $I_\xi$. Therefore we obtain
$$
I_{\bar\rho}=\frac{2}{\sqrt{1-P^2}}\,\Pi\left(\frac{P}{2}\right).
\refstepcounter{suppeq}
\eqno{(\rm S\arabic{suppeq})}
$$

Note that if $S<y$ then $P>1$ independently from $\xi$ and $\theta$, thus $I_{\bar\rho}=0$ and $I_\xi=0$. If $S>y$ then $I_{\bar\rho}\ne0$ only if $|P|<1$. The last condition restricts the limits of integration in $I_\xi$. Taking this into account and integrating over $\xi$ in Eq. (\ref{I_xi}) we obtain
$$
I_\xi=\frac{2\pi}{S^2}(S+y\cos\theta)H(S-y),
\refstepcounter{suppeq}
\eqno{(\rm S\arabic{suppeq})}
\label{I_xi_2}
$$
where $H(x)$ is the Heaviside step function.

Thus, the single-photon PQPD is obtained by taking the derivative of $I_\xi$ in Eq. (\ref{W1_2}),
$$
W_1(S,\theta,\phi)=\frac{p_1\cos\theta}{4\pi S^2}\delta(S-1)-\frac{p_1(1+\cos\theta)}{4\pi S}\delta'(S-1).
\refstepcounter{suppeq}
\eqno{(\rm S\arabic{suppeq})}
\label{W_1_3}
$$
Here we used the fact that $\frac{d}{dx}\left[xH(x)\right]=H(x)$. This expression can be obtained using any approximation for the Heaviside step function. In our opinion this result is more physical than the one obtained by a formal derivation.

The final theoretical distribution (13) is achieved by summing the no-photon $W_0$ (\ref{W_0}) and the single-photon $W_1$ (\ref{W_1_3}) PQPD.